\documentclass[twocolumn,prl,aps]{revtex4}

\usepackage{amssymb}
\usepackage{bm}% bold math
\usepackage{epsfig}
\usepackage{graphicx}
\usepackage{amsmath}

\def\lambdaf{$\lambda_F$}

\def\sq23{23$\times \sqrt{3}$}

\begin{document}

\title{Long-range repulsive interaction between TTF molecules on a metal surface
\\
induced by charge transfer}

\author{I. Fernandez-Torrente$^1$, S. Monturet$^2$, K.J. Franke$^1$, J. Fraxedas$^3$, N. Lorente$^{2,3}$, and J.I. Pascual$^1$}

\affiliation{$^1$ Institut f\"ur Experimentalphysik, Freie
Universit\"at Berlin, Arnimallee 14, 14195 Berlin, Germany\\
$^2$ Laboratoire Collisions, Agr\'egats, R\'eactivit\'e. UMR5589.
Universit\'e Paul Sabatier, 118 route de Narbonne, 31062 Toulouse,
c\'edex France\\
%$^3$ Institut de Ci\`encia de Materials de Barcelona - CSIC. Campus de la
%UAB, 08193 Bellaterra, Spain .
$^3$ Centre d' Investigaci\'o en Nanoci\`encia i Nanotecnologia
(CIN2-CSIC), Edifici CM-7, Campus de la UAB, E-08193 Bellaterra, Spain. }
\date{\today}

\begin{abstract}

The low-coverage adsorption  of a molecular electron donor,
tetrathiafulvalene, on Au(111) is characterized by the spontaneous
formation of superlattice  of monomers, whose spacing exceeds the
equilibrium distance of non-covalent interactions and depends on coverage.
The origin of this peculiar growth mode is due to a long-range repulsive
interaction between molecules. The analysis of molecular-pair
distributions obtained by scanning tunneling microscopy measurements
permits us to determine that the  nature of TTF intermolecular
interactions on Au (111) is electrostatic. A repulsion between molecules
is caused by the accumulation of charge  due to electron donation into the
metal surface, as pictured through density functional theory calculations.
\end{abstract}

\maketitle

The spontaneous formation of self-organized molecular structures at metal
surfaces follows a complex balance of interactions between the basic
functional units \cite{WhiteSci99}. Attractive short-range forces  between
molecules are ubiquitous during growth, but their strength and relevance
varies  depending on the molecular functionalization. For the case of
adsorption on metal surfaces these forces compete with substrate-mediated
interactions, for example, through elastic stress fields
\cite{LauSS77,BruneNAT98,FigeraNAT99} or through surface state electrons
\cite{LauSS78,HyldgaardJCM00,ReppPRL00,KnorrPRB02,WollPRL02,SillyPRL04}.
These usually extend for larger length scales than intermolecular
dispersion forces and can lead to characteristic quasi-periodic arrays of
particles \cite{KnorrPRB02,SillyPRL04}. Long-range interactions can also
have a repulsive nature. This is the case of electrostatic interactions
between charged particles weakly interacting with a non-conducting host
support \cite{WierschemJMP06,SterrerPRL07}, or in  ensembles  of organic
molecules with large dipolar moments on metal
surfaces~\cite{BaberJACS07,Yokoyama2007}.

Apolar and neutral molecules are not expected to build up long-range
interaction potentials other than those mediated by the underlying
substrate \cite{WollPRL02}  and, in most cases, attractive dispersion
forces lead to nucleation in two or three dimensional condensates. Charge
redistribution upon molecular  chemisorption is also able of rendering
interesting changes in the interaction potentials between molecules
\cite{SykesJACS05}. Although this effect is presumably strong in charge
transfer adsorbate systems, it has been usually neglected due to the
screening nature of metallic substrates. An experimental  proof of its
relevance in intermolecular interactions is thus still missing. This could
also help to build up a quantitative picture about fundamental processes
related to molecular charging on metal surfaces.

Here,  we report the  spontaneous formation of quasiperiodic superlattices
of single tetrathiafulvalene (TTF) molecules on a Au(111) surface driven
by local charges at the interface induced upon chemisorption. TTF is well
known as a prototype donor molecule in charge transfer compounds
\cite{TTFTCNQreview}. The free molecule has no electrical dipole moment.
However, on Au(111), it becomes charged upon electron donation. Using a
combination of low-temperature scanning tunneling microscopy and density
functional theory (DFT) we resolve that a repulsive long-range interaction
between charged molecules is built up, thus hindering nucleation in
islands. Through the analysis of molecular pair distributions we
reconstruct the coverage-dependent intermolecular potential wells forming
the molecular lattice.

The experiments were carried out in a custom-made low temperature STM
under ultra-high vacuum.  An atomically clean Au(111) substrate is exposed
to a continuous flux of TTF molecules sublimated from a home-made Knudsen
cell.  TTF has a very high vapor pressure. To obtain the low coverages
used here the crucible was heated to 30$^\circ$C. The sample temperature
was varied during the dosing between 80 K and 300 K, and posteriorly
cooled down to the operating temperature of the STM (4.8 K) for sample
inspection.

%FIGURA stm, etiqueta: stm
\begin{figure}
\centering
\includegraphics[width=0.7\columnwidth]{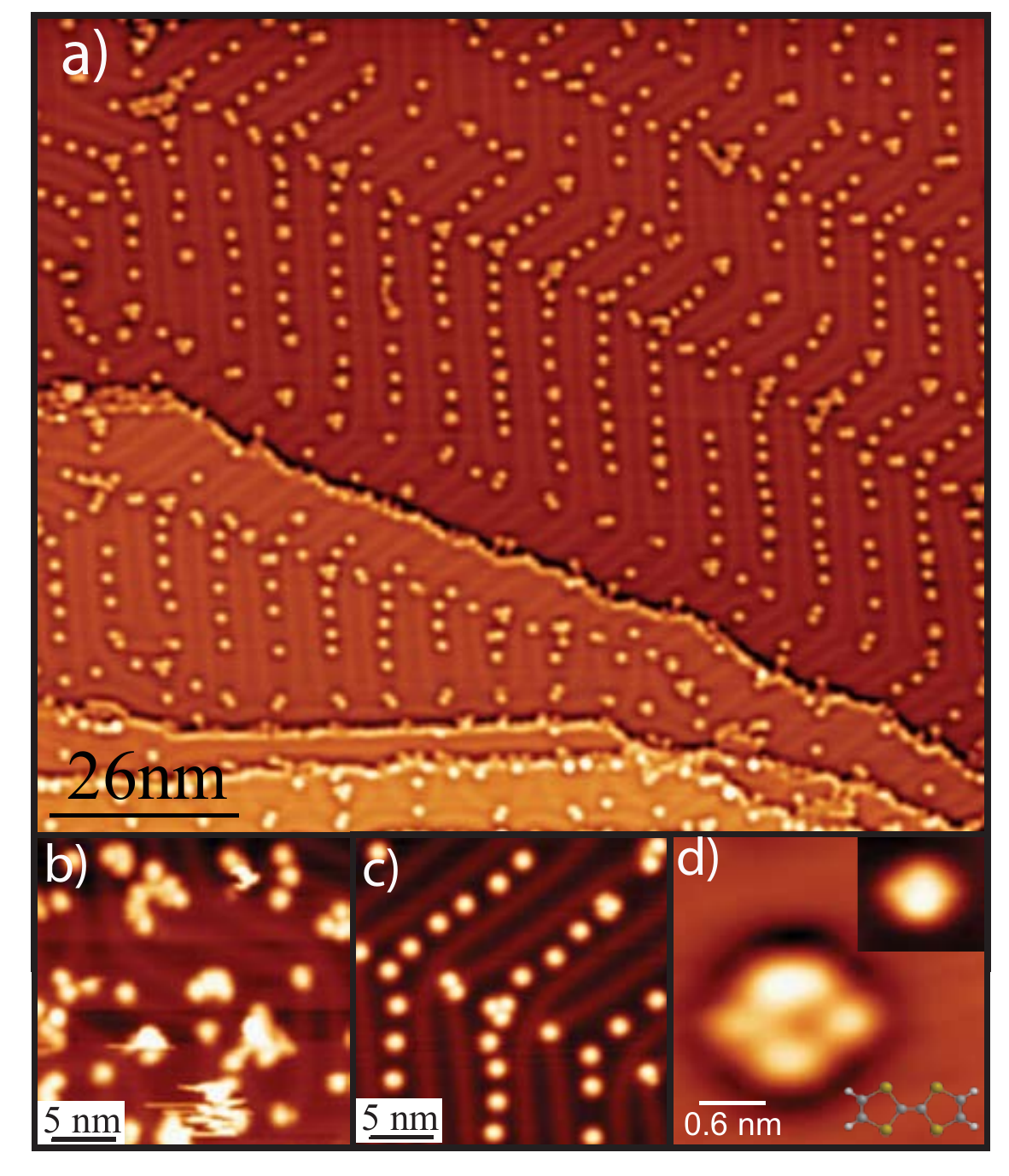}
\caption{ (a) STM image of an Au(111) region with 0.03 ML of TTF deposited
at room temperature. One-dimensional arrays of TTF monomers follow the
Au(111) herringbone reconstruction . (b) Depositing molecules on a cold
sample (80 K) leads to population of a weakly adsorbed precursor state, in
which molecules may nucleate in clusters. After annealing to room
temperature (c), the TTF arrays along FCC regions are formed. (d) STM
image (inset; V$_s$=-1 V) and its Laplace filtered image \cite{wsxm} of a
TTF molecule. The latter reveals that two of the sulfur atoms appear
brighter suggesting a small tilt of the molecular plane with respect to
the surface (later confirmed by theory).} \label{stm}
\end{figure}

Room temperature deposition of a small amount of TTF  ($<0.1$ ML) leads to
the formation of a characteristic quasi-periodic one-dimensional array of
TTF monomers along the FCC regions of the Au(111) 23$\times \sqrt{3}$
reconstruction (Fig.~\ref{stm}). The separation between monomers amounts
several nanometers, $\sim$ 3 nm for the data in Fig.~\ref{stm}(a). This
distance is significantly larger than the typical length scale of
attractive non-covalent interactions. The formation of the superlattice of
TTF monomers needs to be thermally activated. Fig. 1(b) shows the result
of dosing TTF on a 80 K cold sample. In this case both monomers and small
TTF clusters appear randomly spread and are easily dragged by the STM tip,
probably because they populate a weakly adsorbed precursor state. Only
upon annealing the molecules self-organize forming the distinctive
quasi-periodic array, as shown in Fig.~\ref{stm}(c). In this case,
high-resolution STM images (Fig.~\ref{stm}(d)) of intramolecular structure
can be obtained. At negative sample bias we find TTF monomers as composed
by four protrusions. The two largest correspond to the 4 sulfur atoms. The
other two are fainter tails due to the ethylene ends. TTF appears with a
characteristic asymmetry in the images, resembling two of the lateral S
atoms being higher than the other two.

\begin{figure}
\centering
\includegraphics[width=0.7\columnwidth]{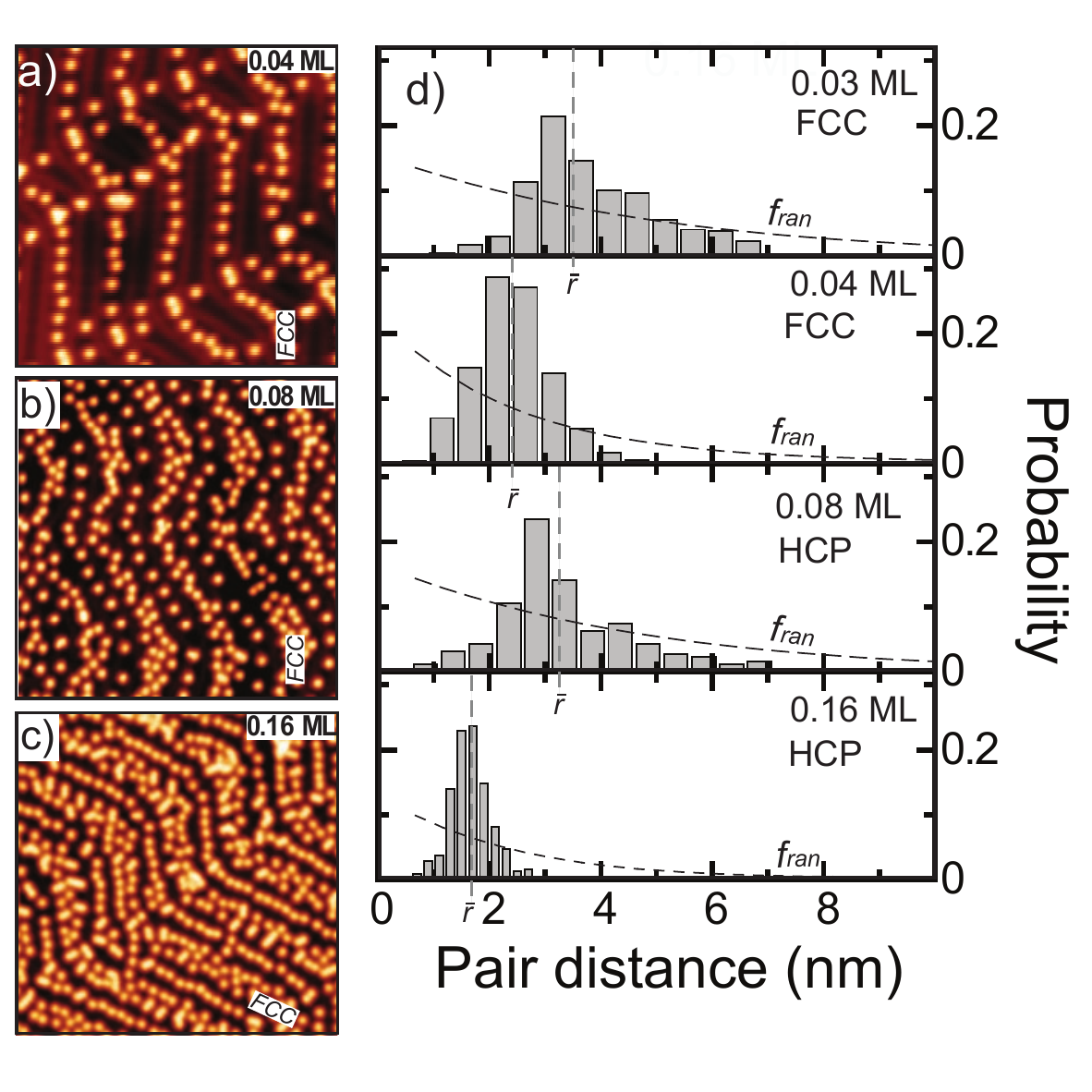}
\caption{ (a-c) STM images of TTF on Au(111) at various coverages. At 0.08
ML molecules appear already at the HCP regions. (d) Pair distributions $f$
of the one-dimensional TTF arrays for the data shown in Fig. 1 (0.03 ML)
and Fig. 2(a-c). For 0.08 ML and 0.16 ML the distributions are performed
on HCP regions. More than 500 pairs are analyzed in each plot. The
molecular coverage is determined from STM images of large surface areas,
assuming that 1 ML corresponds to 2 molecules/nm$^2$. From the lowest to
the largest coverage we obtain  an \textit{average} pair distance
$\overline{r}$ of 3.5 nm, 2.5 nm, 3.3 nm, and 1.7 nm in the one
dimensional arrays. The corresponding 1D distribution functions for non
interacting particles $f_{ran}$ are included. } \label{dist}
\end{figure}

Our results clearly indicate that, in the low-coverage limit and after
annealing, TTF does not respond to attractive forces like, for example,
hydrogen bonding to sulfur atoms \cite{WennJCP03}, avoiding nucleation
into islands. Such behavior prevails as the coverage is increased,
accompanied by a monotonous decrease in the average pair distance
(Fig.~\ref{dist}(a-c)). At 0.08 ML the array is compressed (average pair
distance $\sim 2 nm$) into double rows of monomers   in the FCC regions of
the reconstruction. Close to this coverage HCP regions start also to be
populated with similar one-dimensional  arrays of TTF monomers. Such
tendency to avoid nucleation through the formation of quasi-periodic
molecular arrays is indicative of a long-range interaction mechanism
different from (shorter-range) non-covalent dispersion forces between
molecules.

%FIGURA de distribucion con etiqueta: dist

Elastic deformation of the substrate can lead to long-range interactions
between adsorbates~\cite{LauSS78,BruneNAT98,FigeraNAT99}. The induced
stress field can oppose the approach of two adsorbates becoming the
driving force of an ordered phase. Indeed, the Au (111) herringbone
reconstruction is itself a stressed atomic layer, and therefore, the
periodicity of its folding is very sensitive to small changes in the
stored elastic energy. Our data show that the herringbone structure is
unaffected by a sub-monolayer coverage of TTF. Therefore, this mechanism
is improbable in our case. An alternative mechanism for long-range
interaction between atoms \cite{ReppPRL00,KnorrPRB02} or molecules
\cite{WollPRL02} on metal surfaces  is the interaction potential
associated with the Friedel oscillations  due to the scattering of
surface-state electrons with the adsorbates. A key element in this
mechanism is the oscillatory character of the interaction with a period
related to half of the electronic Fermi wavelength (\lambdaf/2). For
Au(111) this corresponds to 1.8 nm, much smaller than the average pair
distance of the data in Fig.~\ref{stm}. Furthermore, the average pair
distance decreases monotonously with increasing  TTF density along the
rows (Fig.~\ref{dist}). Thus, an interaction mediated by surface electrons
can also be discarded as the driving force leading to the superlattice
formation.

Fig.~\ref{dist}(d) shows the  pair distance $r$ distributions of
one-dimensional arrays  for various coverages (along FCC regions or HCP
regions depending on the coverage). For a one-dimensional system of
non-interacting particles the first-neighbors' random pair distribution
function $f_{ran}$ decays monotonically with the pair distance $r$ as
shown by a dashed line in  figure \cite{note}. The peaked distributions
in Fig.~\ref{dist}(d) are symptomatic of a repulsive long range
interaction between monomers.  Motivated by the donor nature of TTF as a
free molecule we have performed \textit{ab-initio} calculations in order
to trace back the nature of the molecule-surface interaction and its
effect in the long range repulsion between TTF monomers.

\begin{figure} [t]
\centering
\includegraphics[width=0.7\columnwidth]{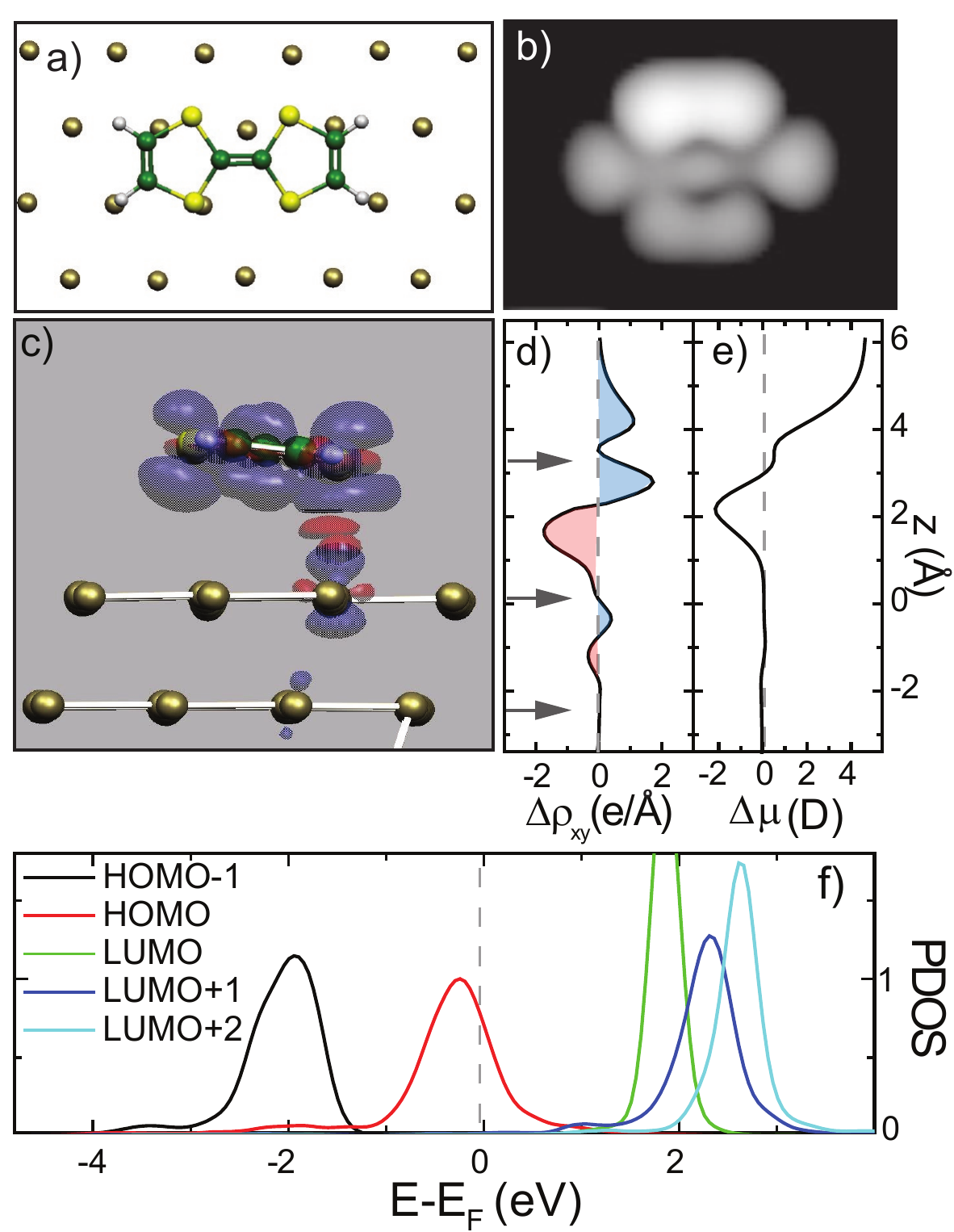}
\caption{ Results from DFT simulations. (a) Fully relaxed configuration of
TTF on Au(111). The uppermost two gold layers as well as the molecular
degrees of freedom are relaxed until atomic forces are lower than 0.01
eV/\AA. (b) Tersoff-Hamman constant current image~\cite{TersoffHamman} of
the molecule in (a) (V=-0.5 V). (c) Induced electronic density by the
molecule--surface interaction. (d) Lateral (x-y planes) integration of the
induced charge. The arrows show the vertical distance values at which the
two topmost surface layers and the two binding S atoms lie. (e)
Accumulated induced dipole. Together with (d) it reveals that the molecule
becomes positively charged. (f) Projected density of states on molecular
orbitals. The electronic states with HOMO character are partially empty,
in agreement with the data of (c-e).} \label{dft}
\end{figure}

We have used density functional theory (DFT) within the generalized
gradient approximation~\cite{PW91} as implemented in the VASP
code~\cite{VASP} to evaluate the properties of a relaxed layer of TTF on
an artificial FCC (111) 4-layers slab of gold atoms. The electron-ion
interaction is described by the projector-augmented wave (PAW)
scheme~\cite{PAW}.  Figure~\ref{dft}(a)  shows the resulting structure of
a relaxed TTF molecule in a 6$\times$4 unit cell. This large unit cell is
employed in order to account for large molecular separations within
computationally reasonable limits.

The interaction of the molecule with the surface is driven by  local S--Au
bonds. Due to the incommensurate dimensions of molecule and surface, the
local interactions lead to an asymmetric chemisorption of the molecule.
%Hence, the molecule rotates and tilts to maximize the number of S--Au
%bonds.
As a result, the molecule aligns along the $\{1\bar{1}0\}$ direction of
the surface and tilts  8$^\circ$ with respect to the surface plane. The
tilt is responsible of the asymmetry  in the experimental constant current
STM image, Fig.~\ref{stm}(d), as it is here captured by its Tersoff-Hamman
simulation~\cite{TersoffHamman} (Fig.~\ref{dft}(b)). At negative bias
voltage, the STM image is basically dominated by the shape of the HOMO.

The local interaction character between Au and S atoms implies a sizable
bonding strength and a large charge donation into the surface. Indeed, the
adsorption energy after dipole corrections is -0.86 eV, and the
surface-molecule distance is 2.76~\AA. The electronic structure of the S
atoms has a large contribution in the highest occupied molecular orbital
(HOMO), what causes a  large redistribution of electronic charge
(Fig.~\ref{dft}(c)). The charge donation is expressed by a partial
decrease of the electron density in the whole molecular plane. The result
is a positive charging of the molecule and the creation of a surplus of
negative charge localized close to the S--Au bonds. Figure~\ref{dft}(d)
shows the planar integration of charge. An excess of positive charge
($\sim 0.6$ $e$) is located about the molecule and the corresponding
screening negative charge ($\sim -0.4$ $e$) is between the molecule and
the first atomic layer. The molecule-surface interaction leads to a large
surface dipole that is evaluated in Fig.~\ref{dft}(e) according to Ref.
\cite{06PRBChen}. The dipole is zero inside the surface and builds up
across the molecule reaching a value of 5.0 D.

The charge donation gleaned from the induced electronic density causes the
partial emptying of the HOMO. This is clearly seen by plotting the
projection of the full electronic structure onto the molecular orbitals
corresponding to the present molecular conformation (Fig.~\ref{dft}(f))
\cite{pdos}. The molecule-surface interaction also broadens the molecular
features associated to the HOMO-1, HOMO and the lowest unoccupied
molecular orbital (LUMO), revealing a substantial hybridization with the
surface electronic structure, while higher-lying orbitals are thinner
showing their small role in the molecular-surface interaction.

\begin{figure} [t]
\centering
\includegraphics[width=0.5\columnwidth]{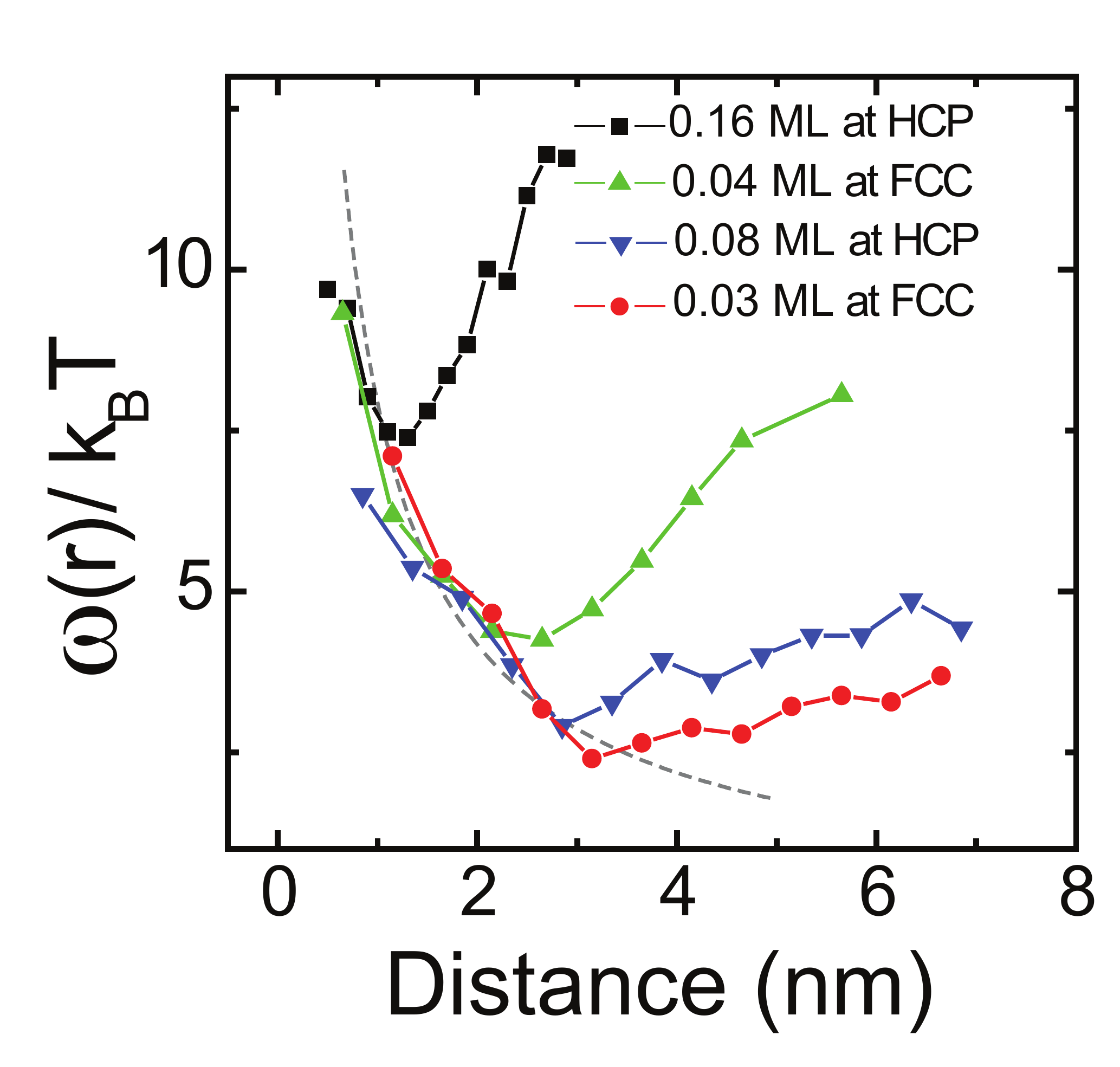}
\caption{ Mean interaction potentials $\omega(r)$ of one-dimensional TTF
arrays obtained from the pair distributions shown in Fig.~\ref{stm}(d).
The dashed line represents the pair electrostatic interaction E(r) between
particles charged with 0.3 $e$ and a temperature (T=160 K) to fit the
repulsive part of $\omega(r)$ for the most dilute case. Each curve has
been shifted upwards an amount (8.4, 5.4, 4.1, 3.8, from top to bottom)
representing the coverage dependent zeroth order internal potential and
approximated here as the electrostatic energy per molecule in a fully
periodic lattice and using the fitted temperature, for consistency. }
\label{fit}
\end{figure}

The ab-initio results evidence a significative charging of the TTF
on the Au(111) surface. For  pair distances shorter than the
Thomas-Fermi screening length on gold surfaces this local charge can
build up a repulsive potential between molecules \cite{ThomasFermi}.
To evidence its role in the formation of the arrays we analyze the
statistics shown in Fig. 2(d). The experimental pair distributions
$f$ arise from the site occupation as dictated by the Boltzmann
factor exp(-($\omega(r)-\mu$)/$k_B$T), where $\omega(r)$ is the
\textit{mean} interaction potential behind the formation of the
superlattice, $\mu$ is a (coverage dependent) zeroth order potential
\cite{note2} and $k_B$ the Boltzmann constant. To evaluate
$\omega(r)$  we divide the experimental pair distribution $f$ by
that of non-interacting particles ($f_{ran}$) and plot
-ln($f/f_{ran}$) (Fig. \ref{fit}). In the limit of a very dilute
system \cite{TTSongPRL73,ReppPRL00,KnorrPRB02}, i.e. where no
quasiperiodic array is formed, $\omega(r)$ would be a good
approximation to the (repulsive) pair interaction potential, E(r).
Here, however, $\omega(r)$ has the shape of a potential well. As the
molecular density increases the well becomes more symmetric and
shallower, in accord with the TTF molecules being confined into
sharper pair distribution and, hence, forming a superlattice.
Unfortunately, it is not trivial to obtain the shape of the pair
interaction E(r) from the mean potential $\omega(r)$
\cite{BernardDiu}. However, we note that for small pair distances
$\omega(r)$ decays as $1/r$ and is consistent with an electrostatic
repulsion between  molecules charged with 0.3 $e$, as it is
described in the \textit{ab initio} results.

Our study has permitted us to show that a highly-ordered chemisorbed phase
of single molecules  can be formed on a metallic surface as a response to
the charge redistribution upon molecular adsorption. Repulsion among
localized charges at the molecule/surface region are strong enough to
hinder nucleation into molecular islands, even when H-bonding between
molecules is expected.  We expect similar behavior to occur in other
molecular systems, thus being a fingerprint of charge transfer processes
at organic/metal interfaces. In this case, the analysis of the
nearest-neighbors pair distribution of molecules at different coverages
turns out to be a very valuable tool for the study of adsorbates'
interaction potential on metallic surfaces.

We acknowledge fruitful discussions with Wolfgang Theis, as well as
financial support of the DAAD and the French Minist\`ere des Affaires
Etrang\`eres (programme PROCOPE), the DFG through  Sfb 658, and
computational resources at the Centre Informatique National de
l'Enseignement Sup\'erieur and the Centre de Calcul Midi-Pyr\'en\'ees.
I.F.T. thanks La Generalitat de Catalunya for her research contract.


\begin{thebibliography}{99}
\bibitem{note} $f_{ran}$ can be exactly obtained from a simple combinatorial
analysis as
$f_{ran}=n!\times(N-r_s)!\times(N-n)!/[(N!\times(N-r_s-n+1)!]$, were $n$
is the number of particles in a box with $N$ sites and $r_s$ is the a pair
distance $r$ in units of sites. This function decreases monotonically with
$r_s$, as a consequence of the one-dimensionality. In this way it differs
from the random pair distribution in two dimensions
\cite{TTSongPRL73,ReppPRL00,KnorrPRB02}
\end{thebibliography}
\end{document}